\documentclass[journal]{IEEEtran}

\usepackage{cite}
\usepackage{citesort}
\usepackage{subfigure}
\usepackage{amsmath}
\usepackage{graphicx}
\usepackage{epstopdf}
\usepackage{paralist}
\ifCLASSINFOpdf

\else
\fi
\begin{document}\fontsize{10}{11.5}\rm
%%%%%%%%%%%%%%%%%%%%%%%%%%%%%%%%%%%%%%%%%%% TITLE %%%%%%%%%%%%%%%%%%%%%%%%%%%%%%%%%%%%%%%%%%
\title{Robust Cooperative Relay Beamforming}

\author{{Behrad Mahboobi, Mehrdad Ardebilipour, Ashkan Kalantari, and Ehsan Soleimani-Nasab}
\thanks{Authors are with the Faculty of Electrical and Computer Engineering, K.N. Toosi University of Technology, P.O. Box 16315-1355, (e-mails: b.mahboobi@ee.kntu.ac.ir, a.kalantari@ee.kntu.ac.ir, ehsan.soleimani@ieee.org, mehrdad@eetd.kntu.ac.ir).}
}
%****************************************************************************
\maketitle

\begin{abstract}
In this paper, the robust distributed relay beamforming problem is solved using the worst case approach, where the problem solution has been involved because of the effect of uncertainty of channel knowledge on the quality of service (QoS) constraints. It is shown that the original robust design, which is a non-convex semi-infinite problem (SIP), can be relaxed and reformed to a semi-definite problem~(SDP). Monte-Carlo simulations are presented to verify the performance improvement of our proposed robust problem over existing robust and non-robust problems in terms of transmit power and symbol error probability.
\end{abstract}

%worst-case optimization
%%%%%%%%%%%%%%%%%%%%%%%%%%%%%%%%%%%%%%%%%%%%%%%%%%%%%%%%%%%% SECTION %%%%%%%%%%%%%%%%%%%%%%%%%%%%%%%%%%%%%%%%%%%%%%%%%%%%%%%%%
\section{Introduction}
\label{Sec: Intro}
%%%%%%%%%%%%%%%%%%%%%%%%%%%%%%%%%%%%%%%%%%%%%%%%%%%%%%%%%%%% SECTION %%%%%%%%%%%%%%%%%%%%%%%%%%%%%%%%%%%%%%%%%%%%%%%%%%%%%%%%%
%Relay networks have been an interesting topic for researchers in recent years because of being a promising
%high-performance capability in the future communication systems.
%Relay networks refer to systems or techniques that allow some relay nodes to help other's
%messages to be relayed to the destination. Amplify-and-forward (AF), decode-and-forward
%(DF), compress-and-forward (CF) and coded cooperation (CC) are the most common and
%popular relaying protocols among several types of relaying
%schemes~\cite{Sendonaris:2003:Part:I,Sendonaris:2003:Part:II,Laneman:2004,Dohler}.
%The AF scheme is more appealing due to its simplicity. In AF relaying technique,
%the relays perform linear operations and forward the signal, but do not decode or
%nonlinearly process it like DF or CF scheme. In the regenerative protocols like DF
%and CC, complicated hardware implementations are required to process and send
%the received signal.
%Multi-input multi-output (MIMO) technology is known to increase the capacity of communication systems.
%Hence, for higher bandwidth demands, relay systems equipped by MIMO technology are one of the candidates for next generation wireless networks.
%Due to the computational complexity of other relaying protocols such as decode-and-forward
%(DF), compress-and-forward (CF)
%~\cite{Dohler}, a good candidates is Amplify-and-forward (AF) relay system.
Relay networks is one of the main novel feasible techniques which can increase the capacity of wireless networks by multi-hopping \cite{raeisi2011optimal,raeisi2011near,mahboobi2010qos} or parallel relaying \cite{alikhani2012interference} in regenerative setting or nonregenerative settings\cite{Dohler}.
Recently, distributed relay beamforming has been found to be appealing because of the simplicity of non-regenerative relays hardware and also achieving the favorite diversity order offered by by multiple route dirverity or . In such systems, relay nodes organize a single virtual MIMO node and transmit the linearly beamformed version of their received signal in distributed fashion without communicating to each other.
% have been mostly investigated in the literature~\cite{Kyoung:Transactions:2010,Ferdinand:Commletter:2011,Firag:CommLetter:2011,
%YueRong:Transactions:2011,Zijian:Transactions:2011,Zhang:TVT:2010}.
%Distributed beamforming
%is one of the new techniques, which
%utilizes single or multiple antenna relays as a virtual multiple antenna system to employ the whole system as a super MIMO beamforming system to benefit more diversity and multiplexing orders~\cite{Dohler}.
%As a result, the system will benefit more diversity and multiplexing orders~\cite{Dohler}.
The distributed beamforming systems for single user~\cite{Havary-Nassab:2008} and multi-users~\cite{Fazeli-Dehkordy:2009:Trans,chaliseMIMO},
%, which consists of a source, a destination and several single antenna relaying
%nodes in a dual-hop scheme. It is supposed that only the second-order statistics
%of all channel coefficients are available at the destination.
 minimize the total relays transmit power with signal to interference and noise ratio (SINR) constraints at the destinations.
%for fairness reasons or QoS constraint scenarios
In~\cite{Havary-Nassab:2008,Fazeli-Dehkordy:2009:Trans,chaliseMIMO}, the instantaneous channel state information (CSI) have to be perfectly available at the relays to maintain the instantaneous SINR above a threshold. %The optimization problem of~\cite{Fazeli-Dehkordy:2009:Trans} is shown to be a non-convex quadratically constraint problem, which is relaxed to a semi-definite programming (SDP) problem and solved using the standard interior point methods~\cite{Nemirovski:2001}.
In all of the proposed beamforming systems in~\cite{Fazeli-Dehkordy:2009:Trans,Havary-Nassab:2008,chaliseMIMO}, it is assumed that the perfect CSI is available at the relay nodes; However, this is an idealistic assumption, since the CSI is often subject to uncertainties because of the channel estimation or quantization error.
%A robust optimization approach that takes into account the CSI uncertainties
%can be employed to consider resource allocation of imperfect CSI systems.
If the statistical information of channels uncertainty (i.e. the probability density function) are available, a probabilistic or statistical approach can be used which the SINR constraints of the problem are often formulated based on outage probability and it is recently investigated in \cite{probabilist_robust2011,probabilist_robust2012}.
In the other case which the unknown perturbation is subject to unknown probability distribution with bounded variation, worst-case robust approach is used commonly. The worst case robust beamforming has been well presented in \cite{boyd_worstcase2008} for basic multiple antenna system with only simple power constraints.
%Bounded uncertainty means that the error is bounded by a specific value
%which the term deterministic uncertainty is used for it in the literature.
%Quantization noise is an example of deterministic ambiguity of CSI.
%Another type of robust optimization method is to use the stochastic method where the
%probability distribution of perturbation is known. The channel estimation error
%which results in the probabilistic outage, is an instance of stochastic
%perturbation and is usually modeled by normal distribution. In such paradigms,
%maximum probability of outage is subject to a predefined value, which is the
%stochastic QoS constraint in the power optimization problem~\cite{chaliseshahbaz}
%,~\cite{Zijian:Transactions:2011}. Some works
%have considered this robust design approach in broadcast beamforming of
%downlink transmission~\cite{Wajid:2009}.
The robust formulation in \cite{chaliseMIMO1} and \cite{Chalise:2010:Trans} has redesigned respectively the distributed beamforming problems of \cite{Fazeli-Dehkordy:2009:Trans} and \cite{chaliseMIMO} by worst-case approach, which demonstrates visible performance improvement with respect to the non-robust systems when the channels are perturbed.\\%of~\cite{Fazeli-Dehkordy:2009:Trans,chaliseMIMO}, their problem is not designed completely robust against all channel uncertainties because the source to relays channels have not been modeled to be subject to uncertainty, mostly because of the mathematical intractability of the problem formulation that they had predicted. No work has designed a robust distributed rely beamforming by perturbing all of dual hop channels. The contributions of our works with respect to earlier works are:\\ %briefly listed as\\%\begin{enumerate}%\item
The contributions of our work are as follows\\
%\begin{itemize}
\begin{compactitem}
 \item All channels are subject to uncertainty, while in \cite{Chalise:2010:Trans,chaliseMIMO1} the sources-to-relays channels are not perturbed.\\
  %\item
 \item The approach in \cite{Chalise:2010:Trans,chaliseMIMO1} used a conservative approximation for SINR constraints, Min over Max (MoM), to avoid semi-infinite programming (SIP) appeared due to the uncertainty region in the QoS constraints of the robust problem. But because of more accurate formulation, our work doesn't utilize conservative MoM approximation and it outperforms \cite{Chalise:2010:Trans,chaliseMIMO1}. Instead of MoM, we propose a new equivalent Semi-Definite representable (SDr) problem to the original SIP.
\end{compactitem}
%    \vspace{-2 cm}
\section{Robust Design Formulation} \label{Sec: Robust Problem}
%\vspace{-.7 cm}
%%%%%%%%%%%%%%%%%%%%%%%%%%%%%%%%%%%%%%%%%%%%%%%%%%%%%%%%%%%% SECTION %%%%%%%%%%%%%%%%%%%%%%%%%%%%%%%%%%%%%%%%%%%%%%%%%%%%%%%%%
%In this Section, the robust counterpart of the non-robust problem of~
%\cite{Fazeli-Dehkordy:2009:Trans} is introduced.
%which was reconsidered in Section~\ref{Sec: Non-robust Problem}.
Assume $d$ single antenna sources and destinations are communicating without direct link through $R$  single antenna amplify-and-forward (AF) relays. The one-shot received  signals at the relays in the scalar and vector forms are written as
 %\vspace{-.5cm}
\begin{align}
{x_r} = \sum\limits_{p = 1}^d {{f_{rp}}{s_p} + {v_r}},\quad
%\label{eqn: Rcd Sig Relay}
{\bf{x}} = \sum\limits_{p = 1}^d {{{\bf{f}}_p}{s_p}}  + {\bf{v}}.
\label{eqn:Rece Sig Relay Vec}
\end{align}
%\vspace{-.9cm}
where $s_p$ is $p^\textrm{th}$ user's transmit signal with the transmit power $P_p=E\left| s_p \right|^2$, $x_r$  and ${v_r}$ are $r^\textrm{th}$ relay received signal and noise, $f_{rp}$ is  the complex channel coefficient from $p^\textrm{th}$ source to $r^\textrm{th}$ relay. For ease of vector based formulation, we define
%\vspace{-.5cm}
\begin{align}
{{\bf{f}}_p} \buildrel \Delta \over = {\left[ {{f_{1,p}}...,{f_{R,p}}} \right]^T}
,{\bf{x}} \buildrel \Delta \over = {\left[ {x_1...,x_R} \right]^T}
, \textbf{v} \buildrel \Delta \over = {\left[ {v_1...,v_R} \right]^T}.
\label{eqn: Channel Vect}
\end{align}
To perform relay beamforming, a complex weight coefficient, denoted as $w_r^*$, is used at the ${r^{th}}$ relay to amplify its received signal. By denoting $\textbf{w} \buildrel \Delta \over={{{\left[ {{w_1},{w_2},...,{w_R}} \right]}^T}}$, $\textbf{W} \buildrel \Delta \over = diag\left( \textbf{w} \right)$, the output signal vector of the relays is $\textbf{t} = {\textbf{W}^H}\textbf{x}$ and the received signal at the $k^{th}$ destination is given by
%\vspace{-.5cm}
\begin{align}
\!\!\!%\label{eqn: Rece Sig Des Vec}
{y_k} =& \textbf{g}_k^T{\bf{t}} + {n_k}
=  \textbf{g}_k^T{{\bf{W}}^H}\sum\limits_{p = 1}^d {{{\bf{f}}_p}} {s_p} + {\bf{g}}_k^T{{\bf{W}}^H}{\bf{v}} + {n_k}
\nonumber \\
=& {\bf{g}}_k^T{{\bf{W}}^H}{{\bf{f}}_k}{s_k}\! +\! {\bf{g}}_k^T{{\bf{W}}^H}\!\!\!\!\!\sum\limits_{p = 1, p \ne k\hfill}^d \!\!\!\!{\!{{\bf{f}}_p}} \!{s_p}\! + \!\left( {{\bf{g}}_k^T{{\bf{W}}^H}{\bf{v}} \!+ \!{n_k}\!} \right).
\label{eqn: Rece Sig kth Des}
\end{align}
where $g_{rk}$ is the channel coefficient from $r^\mathrm{th}$ relay to $k^\mathrm{th}$ destination, $n_k$ is the noise at the $k^\mathrm{th}$ destination and $\textbf{g}_k \buildrel \Delta \over = {\left[ {{g_{1k}}...,{g_{Rk}}} \right]^T}$.
The three last terms in \eqref{eqn: Rece Sig kth Des} are respectively desired received signal, interference and noise.
By denoting $P_T$ as the total transmit relays power, $\gamma_{th}$ as the specified SINR
threshold, $\Gamma_k$ as the SINR at the $k^{th}$ destination and $P_s^k$, $P_i^k$ and $P_n^k$  respectively as the desired signal, interference
and noise powers at the $k^{th}$ destination, they can be computed as
\begin{align}
\!\!\!\!\!\!\!\!{P_T}=&\mathop E\limits_{\left( {{s_1},..,{s_d}} \right)}\left\{ {{\!{\bf{t}}^H}{\bf{t}}} \right\}
\!\!=\!\!Tr\!\left\{ {{{\bf{W}}^H}\mathop E\limits_{\left( {{s_1},..,{s_d}} \right)}\left\{ {{{\bf{x}}^H}{\bf{x}}} \right\}{\bf{W}}} \right\}\nonumber\\
=&Tr\left\{ {{{\bf{W}}^H}{{\bf{R}}_x}{\bf{W}}} \right\}=\sum\limits_{r = 1}^R {{{\left| {{w_r}}\right|}^2}{{\left[ {{{\bf{R}}_x}} \right]}_{r,r}}}={{\bf{w}}^H}{\bf{Dw}}
\label{eqn: P_T Reform}\\
P_s^k=&\mathop E\limits_{ {{s_k}} } {\left| {\bf{g}}_k^T{{\bf{W}}^H}{{\bf{f}}_k}{s_k} \right|^2} %{\bf{g}}_k^T{{\bf{W}}^H}{{\bf{f}}_k}{\bf{f}}_k^H{\bf{Wg}}_k^*E\left\{ {{{\left| {{s_k}} \right|}^2}} \right\}
={P_k}{{\bf{w}}^H}{{\bf{h}}_k}{\bf{h}}_k^H{\bf{w}}
= {{\bf{w}}^H}{\bf{R}} _h^k {\bf{w}}\nonumber\\
%\label{eqn: Received signal Power}
P_i^k=&\!\!\!\!\!\!\mathop E\limits_{\left( {{s_1},..,{s_d}} \right)}\!\!{\left| {{\bf{g}}_k^T{{\bf{W}}^H}\!\!\!\!\!\sum\limits_{p = 1,p \ne k}^d\!\!\!{{\bf{f}}_p}{s_p}} \right|^2}
\!\!\!=\!\!{\left| {\!{P_k}{{\bf{w}}^H}\!\!\left( {{{\bf{g}}_k}\!\odot\!{{\bf{f}}_p}} \right)} \right|^2}
={{\bf{w}}^H}{{\bf{Q}}_k}{\bf{w}}\nonumber\\
%\label{eqn: Interference Power}
P_n^k=&\mathop E\limits_{{\bf{v}},{n_k}} {\left| {{\bf{g}}_k^T{{\bf{W}}^H}{\bf{v}} + {n_k}} \right|^2}
={{\bf{w}}^H}{{\bf{D}}_k}{\bf{w}} + \sigma _n^2
\nonumber \\
%\label{eqn: Noise Power}
\Gamma_k=&\frac{{P_s^k}}{{P_i^k + P_n^k}}=\frac{{{{\bf{w}}^H}{\bf{R}}_h^k{\bf{w}}}}{{{{\bf{w}}^H}\left( {{{\bf{Q}}_k} + {{\bf{D}}_k}} \right){\bf{w}} + \sigma _n^2}}
\label{eqn: SINR definition}
\end{align}
where in the above formulation we have used the following auxiliary matrix and vector parameters
%\vspace{-.8 cm}
\begin{align}
&{{\bf{R}}_x}&&\buildrel \Delta \over =\sum\limits_{p = 1}^d {{P_p}{\bf{R}}_f^p}  + \sigma _v^2{\bf{I}} ,\, {\bf{R}}_f^p \buildrel \Delta \over = {{{\bf{f}}_p}{\bf{f}}_p^H},
\,{\bf{R}}_g^k \buildrel \Delta \over =  {{{\bf{g}}_k}{\bf{g}}_k^H}
\label{eqn: Correlation Matrix}\\
&{\bf{R}}_h^k&&\buildrel \Delta \over ={P_k} {{{\bf{h}}_k}{\bf{h}}_k^H},
\quad {{\bf{h}}_k} \buildrel \Delta \over ={{\bf{g}}_k} \odot {{\bf{f}}_k}
\label{eqn: h_k R_h_K}\\
&{{\bf{Q}}_k}&&\buildrel \Delta \over ={\sum\limits_{p \in \left\{ {1,...,d} \right\}\backslash \{ k\} }^{} {{P_p}{\bf{h}}_k^p{{\left( {{\bf{h}}_k^p} \right)}^H}} \ } , \quad
{\bf{h}}_k^p \buildrel \Delta \over ={{\bf{g}}_k} \odot {{\bf{f}}_p}
\label{eqn: h_k_p Q_k}\\
&{\bf{D}}&&\buildrel \Delta \over = \rm{diag}\left( {{{\left[ {{{\bf{R}}_x}} \right]}_{1,1}},{{\left[ {{{\bf{R}}_x}} \right]}_{2,2}},\cdots,{{\left[ {{{\bf{R}}_x}} \right]}_{R,R}}} \right)\\
&{{\bf{D}}_k}&&\buildrel \Delta \over = \sigma _{v}^2 \rm{diag}\left( {{{\left[ {{\bf{R}}_g^k} \right]}_{1,1}},{{\left[ {{\bf{R}}_g^k} \right]}_{2,2}},\cdots,{{\left[ {{\bf{R}}_g^k} \right]}_{R,R}}} \right).
\label{eqn: D}
\end{align}
%. In both of the cases, channel matrices are erroneous.
%due to the feedback
%and quantization, the channel matrices are erroneous.
%By appropriately modeling the uncertainty of the
%source-relay and relay-destination channels, we show that the robust design problem can be converted to a
%semi-definite problem by applying a relaxation.
The robust beamforming aims to minimize the total relay transmit power $P_T$ subject to holding the SINR of each user $\Gamma_k$, above a predefined $\gamma_k$, while the state information of $\textbf{f}_k$ and $\textbf{g}_k$ are incomplete. Now the robust problem can be written as
%we can rewrite the
%problem \eqref{eqn: Optimization Problem final} as
%%%%%%%%%%%%%%%%%%%%%%%%%%%%%%%%%%%%%%%%%%%%%%%%%%%%%%%%%%%%% EQUATION %%%%%%%%%%%%%%%%%%%%%%%%%%%%%%%%%%%%%%%%%%%%%%%%%%%%%%%%%
%\vspace{-.71 cm}
\begin{subequations}\label{eqn:original_SIP}
\begin{align}
&\mathop {{\rm{minimize}}}\limits_w \,\left\{\mathop{\rm{max}}\,{\hat P_T}\right\}=\left\{\mathop{\rm{max}}\,{\textbf{w}^H}  \widehat{\textbf{D}}\textbf{w}\right\}
 \\
&{\rm{s.t.}}\,\, \Gamma_k=\frac{{{\textbf{w}^H} \widehat{\textbf{R}}_h^k\textbf{w}}}{{{\textbf{w}^H}\left( {{{\widehat{\textbf{Q}}}_k} + {{\widehat{\textbf{D}} }_k}} \right)\textbf{w} + \sigma _n^2}} \ge {\gamma _k}
\, \forall \,k{\rm{ = 1}}...,{\rm{d}}\label{eqn: Perturbed Optimization Problem} \\
&\qquad\forall \,{\widehat{\textbf{Q}}_k} \in {\textbf{S}^{{{\widehat {\textbf{Q}}}_k}}},\,\forall \,{\widehat{\textbf{D}}_k} \in {\textbf{S}^{{{\widehat{\textbf{D}}}_k}}},\,{\widehat{\textbf{D}}} \in {\textbf{S}^{{{\widehat{\textbf{D}}}}}},\,\forall \,{\widehat{\textbf{R}}_k} \in {\textbf{S}^{{{\widehat {\textbf{R}}}_k}}}\,\,.
\label{orignal_perturb_set}
\end{align}
\end{subequations}
%\vspace{-3 mm}
%%%%%%%%%%%%%%%%%%%%%%%%%%%%%%%%%%%%%%%%%%%%%%%%%%%%%%%%%%%%%%%%%%%%%%%%%%%%%%%%%%%%%%%%%%%%%%%%%%%%%%%%%%%%%%%%%%%%%%%%%%%%%%%%
where $\widehat {\bf{D}} = {\bf{D}} + {{\bf{\Delta }}_{\bf{D}}}$, $\widehat {\bf{R}}_h^k = {\bf{R}}_h^k + {\bf{\Delta }}_k^{{R^h}}$, ${\widehat {\bf{Q}}_k} = {{\bf{Q}}_k} + {\bf{\Delta }}_k^{\bf{Q}},\,{\widehat {\bf{D}}_k} = \,{{\bf{D}}_k} + {\bf{\Delta }}_k^{\bf{D}}$ are the perturbed versions of the matrices defined in \eqref{eqn: Correlation Matrix}-\eqref{eqn: D}. The block perturbation of the matrices is an approach to simplify the algebraic complexity appears due to propagation of perturbation from $\textbf{f}_k$ and $\textbf{g}_k$ to the matrices defined in \eqref{eqn: Correlation Matrix}-\eqref{eqn: D}. This approach is proposed by ~\cite{Chalise:2010:Trans,chaliseMIMO1}. In their works $\textbf{f}_k$ is left unperturbed.
%\vspace{-.8 cm}
%\begin{align}
%&{{\bf{R}}_x} && \buildrel \Delta \over = \sum\limits_{p = 1}^d {{P_p}{\bf{R}}_f^p}  + \sigma _v^2{\bf{I}} ,\, {\bf{R}}_f^p \buildrel \Delta \over = {{{\bf{f}}_p}{\bf{f}}_p^H},
%\,{\bf{R}}_g^k \buildrel \Delta \over =  {{{\bf{g}}_k}{\bf{g}}_k^H}
%\label{eqn: Correlation Matrix}\\
%&{\bf{R}}_h^k && \buildrel \Delta \over = {P_k} {{{\bf{h}}_k}{\bf{h}}_k^H},
%\quad {{\bf{h}}_k} \buildrel \Delta \over = {{\bf{g}}_k} \odot {{\bf{f}}_k}
%\label{eqn: h_k R_h_K}\\
%&{{\bf{Q}}_k} &&\buildrel \Delta \over =  {\sum\limits_{p \in {D_k}}^{} {{P_p}{\bf{h}}_k^p{{\left( {{\bf{h}}_k^p} \right)}^H}} } , \quad
%{\bf{h}}_k^p \buildrel \Delta \over = {{\bf{g}}_k} \odot {{\bf{f}}_p}
%\label{eqn: h_k_p Q_k}\\
%&{\bf{D}} && \buildrel \Delta \over = diag\left( {{{\left[ {{{\bf{R}}_x}} \right]}_{1,1}},{{\left[ {{{\bf{R}}_x}} \right]}_{2,2}},...,{{\left[ {{{\bf{R}}_x}} \right]}_{R,R}}} \right)\\
%&{{\bf{D}}_k} &&\buildrel \Delta \over = \sigma _{v}^2 diag\left( {{{\left[ {{\bf{R}}_g^k} \right]}_{1,1}},{{\left[ {{\bf{R}}_g^k} \right]}_{2,2}},...,{{\left[ {{\bf{R}}_g^k} \right]}_{R,R}}} \right)
%\label{eqn: D}
%\end{align}

The matrices ${\bf{\Delta } _{\bf{D}}}$, ${\bf{\Delta}} _k^{{\textbf{R}^h}}$, ${\bf{\Delta }} _k^\textbf{Q}$ and ${\bf{\Delta }} _k^{\bf{D}}$ are the random uncertainty Hermitian matrices which are added to $\textbf{D}$, $\textbf{R}_k^h$, $\textbf{Q}_k$ and $\textbf{D}_k$, respectively. The sets of all possible values of $\widehat {\bf{D}}$, $\widehat {\bf{R}}_h^k$, ${\widehat {\bf{Q}}_k},\,{\widehat {\bf{D}}_k}$ are denoted by ${{\bf{S}}^{\widehat {\bf{D}}}}$, ${{\bf{S}}^{\widehat {\bf{R}}_h^k}}$, ${{\bf{S}}^{{{\widehat {\bf{Q}}}_k}}}$
and ${{\bf{S}}^{{{\widehat {\bf{D}}}_k}}}$ respectively, which cover all cases of the perturbed channel
coefficients. It is assumed that the ${\bf{\Delta} _{\bf{D}}}$ and $\bf{\Delta} _k^{\bf{D}}$ are diagonal random
matrices, because they are perturbing $\textbf{D}$ and $\textbf{D}_k$ matrices which are also
diagonal. Furthermore, according to the worst-case approach, we assume that the channel coefficient
uncertainties are norm bounded by some known constants as
%%%%%%%%%%%%%%%%%%%%%%%%%%%%%%%%%%%%%%%%%%%%%%%%%%%%%%%%%%%%% EQUATION %%%%%%%%%%%%%%%%%%%%%%%%%%%%%%%%%%%%%%%%%%%%%%%%%%%%%%%%%
\begin{subequations}\label{eqn: Perturbation Limit}
\begin{align}
&\left\|{\left. {{{\bf{\Delta }}_{\bf{D}}}} \right\|} \right. \! \le \! {\varepsilon _{\bf{D}}}\label{eqn: Perturbation Limita}\\
&\left\| {\left. {{\bf{\Delta }}_k^\textbf{Q}} \right\|} \right. \le \varepsilon _k^\textbf{Q},\left\| {\left. {{\bf{\Delta }}_k^{\bf{D}}} \right\|} \right. \le \varepsilon _k^{\bf{D}},\,\left\| {\left. {{\bf{\Delta }}_k^{{\textbf{R}^h}}} \right\|} \right. \le \varepsilon _k^{{\textbf{R}^h}}\label{eqn: Perturbation Limitb}.
\end{align}
\end{subequations}
%%%%%%%%%%%%%%%%%%%%%%%%%%%%%%%%%%%%%%%%%%%%%%%%%%%%%%%%%%%%%%%%%%%%%%%%%%%%%%%%%%%%%%%%%%%%%%%%%%%%%%%%%%%%%%%%%%%%%%%%%%%%%%%%
Since the perturbation matrices are not independent in general, the problem formulated by replacing \eqref{orignal_perturb_set} by \eqref{eqn: Perturbation Limit}, results in suboptimal solution for \eqref{eqn:original_SIP}.
In order to guarantee positive power quantities and the convexity of our problem, the estimated channel matrices should be positive semi-definite as
%%%%%%%%%%%%%%%%%%%%%%%%%%%%%%%%%%%%%%%%%%%%%%%%%%%%%%%%%%%%% EQUATION %%%%%%%%%%%%%%%%%%%%%%%%%%%%%%%%%%%%%%%%%%%%%%%%%%%%%%%%%
\begin{subequations}\label{eqn: PSD Condition}
\begin{align}
&{\bf{D}} + {{\bf{\Delta }}_{\bf{D}}}\underline\succ0\label{eqn: PSD Conditiona}\\
&{\bf{R}}_h^k + {\bf{\Delta }}_k^{{{\bf{R}}^h}}\underline\succ0,{{\bf{Q}}_k} + {\bf{\Delta }}_k^{\bf{Q}}{\rm{ }}\underline\succ0,{{\bf{D}}_k} + {\bf{\Delta }}_k^{\bf{D}}\underline\succ0\label{eqn: PSD Conditionb}.
\end{align}
\end{subequations}
%%%%%%%%%%%%%%%%%%%%%%%%%%%%%%%%%%%%%%%%%%%%%%%%%%%%%%%%%%%%%%%%%%%%%%%%%%%%%%%%%%%%%%%%%%%%%%%%%%%%%%%%%%%%%%%%%%%%%%%%%%%%%%%%
The first intractability of the robust optimization problem in \eqref{eqn: Perturbed Optimization Problem} is
the infinite number of the constraints which the problem should be solved subject to them. In fact, the problem
is a SIP \cite{hu1989semi}. Since SIP have some solutions \cite{hu1989semi}, but computationally complex, it is preferred to avoid these problems by converting them to another standard form. To simplify the SIP robust problem~\eqref{eqn: Perturbed Optimization Problem}, first we need to maximize
the objective function over the uncertainty matrix ${{\bf{\Delta }}_{\bf{D}}}$. Then, the objective function
for the robust problem is reformulated as
%%%%%%%%%%%%%%%%%%%%%%%%%%%%%%%%%%%%%%%%%%%%%%%%%%%%%%%%%%%%% EQUATION %%%%%%%%%%%%%%%%%%%%%%%%%%%%%%%%%%%%%%%%%%%%%%%%%%%%%%%%%
\vspace{-.11 cm}
\begin{align}
{ {\mkern 1mu} {\mkern 1mu} {\mkern 1mu} \mathop {\max }\limits_{{{\bf{\Delta }}_{\bf{D}}}} \,\,{\mkern 1mu} \{ {\mkern 1mu} {{\bf{w}}^H}\left( {{\bf{D}} + {{\bf{\Delta }}_{\bf{D}}}} \right){\bf{w}}\} }
\label{eqn: Objective Function Robust}
\end{align}
%%%%%%%%%%%%%%%%%%%%%%%%%%%%%%%%%%%%%%%%%%%%%%%%%%%%%%%%%%%%%%%%%%%%%%%%%%%%%%%%%%%%%%%%%%%%%%%%%%%%%%%%%%%%%%%%%%%%%%%%%%%%%%%%
Second, since all constraints of \eqref{eqn: Perturbed Optimization Problem} must be satisfied for all values of perturbed matrices, we can equivalently say that minimum value of the left side of the inequality should always be greater than the requested SINR. Considering all of the constraints, the robust optimization problem in \eqref{eqn: Perturbed Optimization Problem}
can be expressed as
%%%%%%%%%%%%%%%%%%%%%%%%%%%%%%%%%%%%%%%%%%%%%%%%%%%%%%%%%%%%% EQUATION %%%%%%%%%%%%%%%%%%%%%%%%%%%%%%%%%%%%%%%%%%%%%%%%%%%%%%%%%
\begin{subequations}\label{eqn: Robust Optimization Problem}
\begin{align}
\mathop {{\rm{Minimize}}}\limits_\textbf{w} \,\,\,\mathop {\max }\limits_{{\bf{\Delta _D}}}& {\mkern 1mu} {\mkern 1mu} {\mkern 1mu} {\textbf{w}^H}\left( {\textbf{D} + {\bf{\Delta _D}}} \right)\textbf{w}
 \\
{\rm{s.}}\,{\rm{t.}}\mathop {\min }\limits_{{\bf{\Delta}} _k^{\textbf{D}},{\bf{\Delta}} _k^{{\textbf{R}^h}},{\bf{\Delta}} _k^\textbf{Q}}& \left\{ {{\textbf{w}^H}\left( {{\textbf{T}_k} + {\bf{\Delta}} _k^\textbf{T}} \right){\textbf{w}_k}} \right\} \ge \sigma _n^2\gamma_k,\forall k
\label{constraintminmax}\\
 \textrm{Inequalities of} \quad &\eqref{eqn: Perturbation Limit} , \eqref{eqn: PSD Condition}\,\forall k=1,\cdots,d.
%& \hspace{-2mm} \textbf{D} + {{\bf{\Delta}} _\textbf{D}}\succeq0, {\mkern 1mu} \,\,{\mkern 1mu} \,\textbf{R}_h^k + {\bf{\Delta}} _k^{{\textbf{R}^h}}\succeq0,\,\,{\mkern 1mu} {\mkern 1mu} {\textbf{Q}_k} + {\bf{\Delta}} _k^\textbf{Q}\succeq0,{\mkern 1mu} \,{\mkern 1mu} {\textbf{D}_k} + {\bf{\Delta}} _k^\textbf{D}\succeq0,
%& \hspace{-2mm} \left\| {\left. {{\bf{\Delta} _\textbf{D}}} \right\|} \right. \le {\varepsilon _\textbf{D}},{\mkern 1mu} {\mkern 1mu} {\mkern 1mu} \left\| {\left. {{\bf{\Delta}} _k^\textbf{Q}} \right\|} \right. \le \varepsilon _k^\textbf{Q},{\mkern 1mu} {\mkern 1mu} \left\| {\left. {{\bf{\Delta}} _k^\textbf{D}} \right\|} \right. \le \varepsilon _k^\textbf{D},{\mkern 1mu} {\mkern 1mu} \left\| {\left. {{\bf{\Delta}} _k^{{\textbf{R}^h}}} \right\|} \right. \le \varepsilon _k^{{\textbf{R}^h}}.
\end{align}
\end{subequations}
%%%%%%%%%%%%%%%%%%%%%%%%%%%%%%%%%%%%%%%%%%%%%%%%%%%%%%%%%%%%%%%%%%%%%%%%%%%%%%%%%%%%%%%%%%%%%%%%%%%%%%%%%%%%%%%%%%%%%%%%%%%%%%%%
%By this equivalent form, we can recast the constraint
%in \eqref{eqn: Perturbed Optimization Problem} as
%%%%%%%%%%%%%%%%%%%%%%%%%%%%%%%%%%%%%%%%%%%%%%%%%%%%%%%%%%%%%% EQUATION %%%%%%%%%%%%%%%%%%%%%%%%%%%%%%%%%%%%%%%%%%%%%%%%%%%%%%%%%
%\begin{align}
%{\rm{  }}\mathop {{\rm{min}}}\limits_{{{\widehat {\bf{Q}}}_k},{{\widehat {\bf{D}}}_k},\widehat {\bf{R}}_h^k} \left\{ {\frac{{{{\bf{w}}^H}\widehat {\bf{R}}_h^k{\bf{w}}}}{{{{\bf{w}}^H}\left( {{{\widehat {\bf{Q}}}_k} + {{\widehat {\bf{D}}}_k}} \right){\bf{w}} + \sigma _n^2}}} \right\} \ge {\gamma _k},{\rm{ }} \,\,\, \forall \,\,\, k{\rm{ = 1,2,}}...{\rm{,d}}{\rm{.}}
%\label{eqn: Constraint Robust 1}
%\end{align}
%%%%%%%%%%%%%%%%%%%%%%%%%%%%%%%%%%%%%%%%%%%%%%%%%%%%%%%%%%%%%%%%%%%%%%%%%%%%%%%%%%%%%%%%%%%%%%%%%%%%%%%%%%%%%%%%%%%%%%%%%%%%%%%%%
%or equivalently
%%%%%%%%%%%%%%%%%%%%%%%%%%%%%%%%%%%%%%%%%%%%%%%%%%%%%%%%%%%%%% EQUATION %%%%%%%%%%%%%%%%%%%%%%%%%%%%%%%%%%%%%%%%%%%%%%%%%%%%%%%%%
%\begin{align}
%{\rm{  }}\mathop {{\rm{min}}}\limits_{{\bf{\Delta }}_k^{{\textbf{R}^h}},{\bf{\Delta }}_k^\textbf{Q},{\bf{\Delta }}_k^\textbf{D}} {{\bf{w}}^H}\left( {{{\bf{T}}_k} + {\bf{\Delta }}_k^{\bf{T}}} \right){\bf{w}} \ge {\gamma _k}\sigma _n^2,{\rm{ }}\forall \,\,\,k{\rm{ = 1,2,}}...{\rm{,d}}
%\label{eqn: Constraint Robust 2}
%\end{align}
%%%%%%%%%%%%%%%%%%%%%%%%%%%%%%%%%%%%%%%%%%%%%%%%%%%%%%%%%%%%%%%%%%%%%%%%%%%%%%%%%%%%%%%%%%%%%%%%%%%%%%%%%%%%%%%%%%%%%%%%%%%%%%%%%
where ${{\bf{T}}_k} = {\bf{R}}_k^h - {\gamma _k}\left( {{{\bf{Q}}_k} + {{\bf{D}}_k}} \right)$
and ${\bf{\Delta }}_k^\textbf{T}
= {\bf{\Delta }}_k^{{\textbf{R}^h}}
- {\gamma _{th}}\left( {{\bf{\Delta }}_k^\textbf{Q} + {\bf{\Delta }}_k^\textbf{D}} \right)$.\\
In contrast to our constraint formulation \eqref{constraintminmax},  the approach of \cite{Chalise:2010:Trans,chaliseMIMO1} has differently qualified $\Gamma_k\geq \gamma_k$ by approximating $\frac{\rm{min}{P_s^k}}{\rm{max}\left({P_i^k + P_n^k}\right)}\ge\gamma_k$ which is named MoM in our work.\\
%%%%%%%%%%%%%%%%%%%%%%%%%%%%%%%%%%%%%%%%%%%%%%%%%%%%%%%%%%%%%%%%%%%%%%%%%%%%%%%%%%%%%%%%%%%%%%%%%%%%%%%%%%%%%%%%%%%%%%%%%%%%%%%%%
In order to use the maximum term in the objective function of \eqref{eqn: Robust Optimization Problem}, along with
its active constraint $\textbf{D} + {\bf{\Delta _D}}\succeq0$, we use the Rayleigh-Ritz
theorem. Actually, when ${\bf{\Delta _D}}$ is in the same
direction of $\textbf{w}{\textbf{w}^H}$, the objective function is maximized. The maximization
over ${\bf{\Delta _D}}$ subject to its related
constraints in~\eqref{eqn: Robust Optimization Problem} is
%%%%%%%%%%%%%%%%%%%%%%%%%%%%%%%%%%%%%%%%%%%%%%%%%%%%%%%%%%%%% EQUATION %%%%%%%%%%%%%%%%%%%%%%%%%%%%%%%%%%%%%%%%%%%%%%%%%%%%%%%%%
\begin{align}
&\mathop {{\rm{max.}}}\limits_{{\bf{\Delta _D}}} {\mkern 1mu} {\mkern 1mu} {\mkern 1mu} \,\,{\textbf{w}^H}\left( {\textbf{D} + {\bf{\Delta _D}}} \right)\textbf{w}
\nonumber \\
&{\rm{s.}}\,{\rm{t.}}\,\,\,\,\textbf{D} + {\bf{\Delta _D}}\succeq0,\,\,\left\| {{\bf{\Delta _D}}} \right\| \le {\varepsilon _\textbf{D}}
\label{eqn: Robust Optimization Objective}
\end{align}
%%%%%%%%%%%%%%%%%%%%%%%%%%%%%%%%%%%%%%%%%%%%%%%%%%%%%%%%%%%%%%%%%%%%%%%%%%%%%%%%%%%%%%%%%%%%%%%%%%%%%%%%%%%%%%%%%%%%%%%%%%%%%%%%
where the worst case value for ${\bf{\Delta _D}}$ is
%%%%%%%%%%%%%%%%%%%%%%%%%%%%%%%%%%%%%%%%%%%%%%%%%%%%%%%%%%%%% EQUATION %%%%%%%%%%%%%%%%%%%%%%%%%%%%%%%%%%%%%%%%%%%%%%%%%%%%%%%%%
%\begin{align}
%{{\bf{w}}^H}\left( {{\bf{D}} + {\varepsilon _{\bf{D}}}\frac{{{\bf{w}}{{\bf{w}}^H}}}{{\left\| {\bf{w}} \right\|}^2}} \right){\bf{w}}.
%\label{eqn: Robust Optimization Objective 2}
%\end{align}
\begin{align}
{\bf{\Delta _D}}=\frac{{{\bf{w}}{{\bf{w}}^H}}}{{\left\| {\bf{w}} \right\|}^2}.
\label{eqn: Robust Optimization Objective 2}
\end{align}
%%%%%%%%%%%%%%%%%%%%%%%%%%%%%%%%%%%%%%%%%%%%%%%%%%%%%%%%%%%%%%%%%%%%%%%%%%%%%%%%%%%%%%%%%%%%%%%%%%%%%%%%%%%%%%%%%%%%%%%%%%%%%%%%
Using~\eqref{eqn: Robust Optimization Objective 2}, the robust form of the
objective function of \eqref{eqn: Robust Optimization Problem} will be
%%%%%%%%%%%%%%%%%%%%%%%%%%%%%%%%%%%%%%%%%%%%%%%%%%%%%%%%%%%%% EQUATION %%%%%%%%%%%%%%%%%%%%%%%%%%%%%%%%%%%%%%%%%%%%%%%%%%%%%%%%%
\vspace{-.41 cm}
\begin{align}
%\mathop {{\rm{min.}}}\limits_\textbf{w} \,\,\,\,\,\,\,\,\,
{\mkern 1mu} {\mkern 1mu} {\mkern 1mu} {\textbf{w}^H}\left( {\textbf{D} + {\varepsilon _\textbf{D}}\textbf{I}} \right)\textbf{w}
\label{eqn: Robust Optimization Objective 3}
\end{align}
%%%%%%%%%%%%%%%%%%%%%%%%%%%%%%%%%%%%%%%%%%%%%%%%%%%%%%%%%%%%%%%%%%%%%%%%%%%%%%%%%%%%%%%%%%%%%%%%%%%%%%%%%%%%%%%%%%%%%%%%%%%%%%%%
Since the objective function and QoS constraints of \eqref{eqn: Robust Optimization Problem} do not
have any common constrains, the QoS and positive semi-definite constraints
of \eqref{eqn: Robust Optimization Problem} form an optimization problem over all
of the perturbation matrices, which is independent from the objective function
in \eqref{eqn: Robust Optimization Objective}. Therefore, we focus on the optimization of
the QoS constraint in \eqref{eqn: Robust Optimization Problem} for $k = 1,...,d$. The constraint
in \eqref{eqn: Robust Optimization Problem} can be written as follows
%%%%%%%%%%%%%%%%%%%%%%%%%%%%%%%%%%%%%%%%%%%%%%%%%%%%%%%%%%%%% EQUATION %%%%%%%%%%%%%%%%%%%%%%%%%%%%%%%%%%%%%%%%%%%%%%%%%%%%%%%%%
\begin{IEEEeqnarray}{lCl}
&& \mathop {\rm{min.}}\limits_{ {\bf{\Delta}} _k^\textbf{D}, {\bf{\Delta}} _k^{{\textbf{R}^h}},{\bf{\Delta}} _k^\textbf{Q}} {\textbf{w}^H}\left( {{\textbf{T}_k} + {\bf{\Delta}} _k^\textbf{T}} \right){\textbf{w}_k} - \sigma _n^2\gamma
\nonumber \\
&& {\rm{s.\,t.}} \quad \eqref{eqn: Perturbation Limitb} , \eqref{eqn: PSD Conditionb}\,
%\textbf{R}_h^k + {\bf{\Delta}} _k^{{\textbf{R}^h}}\succeq0,\,\,{\mkern 1mu} {\mkern 1mu} {\textbf{Q}_k} + {\bf{\Delta}} _k^\textbf{Q}\succeq0,{\mkern 1mu} \,{\mkern 1mu} {\textbf{D}_k} + {\bf{\Delta}} _k^\textbf{D}\succeq0
%\nonumber \\
%&&\qquad\left\| {\left. {{\bf{\Delta}} _k^\textbf{Q}} \right\|} \right. \le \varepsilon _k^\textbf{Q},{\mkern 1mu} {\mkern 1mu} \left\| {\left. {{\bf{\Delta}} _k^\textbf{D}} \right\|} \right. \le \varepsilon _k^\textbf{D},{\mkern 1mu} {\mkern 1mu} \left\| {\left. {{\bf{\Delta}} _k^{{\textbf{R}^h}}} \right\|} \right. \le \varepsilon _k^{{\textbf{R}^h}}
\forall \,\,\,\,k = 1,...,d \,,
\label{eqn: Robust Optimization Constraints}
\end{IEEEeqnarray}
%%%%%%%%%%%%%%%%%%%%%%%%%%%%%%%%%%%%%%%%%%%%%%%%%%%%%%%%%%%%%%%%%%%%%%%%%%%%%%%%%%%%%%%%%%%%%%%%%%%%%%%%%%%%%%%%%%%%%%%%%%%%%%%%
Note that the positive semi-definite (PSD) constraint of \eqref{eqn: Robust Optimization Constraints} also satisfies
the corresponding constraints on its related instantaneous covariance matrices, $\textbf{R}_h^k$, ${\textbf{Q}_k}$
and ${\textbf{D}_k}$.
%Also, note that the current form of the formula \eqref{eqn: Robust Optimization Constraints} is
%a non-convex problem similar to the original non-robust problem discussed in the previous section.
To solve the problem, we look for a relaxation scheme to convert the problem into a convex form and investigate the
gap between the relaxed and non-relaxed problems. Applying Lagrange duality technique, the solution of \eqref{eqn: Robust Optimization Constraints} is equal to the solution of the following problem
%%%%%%%%%%%%%%%%%%%%%%%%%%%%%%%%%%%%%%%%%%%%%%%%%%%%%%%%%%%%% EQUATION %%%%%%%%%%%%%%%%%%%%%%%%%%%%%%%%%%%%%%%%%%%%%%%%%%%%%%%%%
\begin{align}
\mathop {\inf }\limits_{{\bf{\Delta}} _k^\textbf{D},{\bf{\Delta}} _k^{{\textbf{R}^h}},{\bf{\Delta}} _k^\textbf{Q}}& \,\,L\left( {\lambda _k^\textbf{Q},\lambda _k^\textbf{D},\lambda _k^{{\textbf{R}^h}},{\bf{Z}}_k^\textbf{Q},{\bf{Z}}_k^\textbf{D},{\bf{Z}}_k^{{\textbf{R}^h}},{\bf{\Delta }}_k^\textbf{Q},{\bf{\Delta }}_k^\textbf{D},{\bf{\Delta }}_k^{{\textbf{R}^h}}} \right)\,
\nonumber \\
\rm{s.\,t.}& \,\,\,  {\bf{Z}}_k^\textbf{Q}{\rm{ }}\underline  \succ  {\rm{ }}0,{\rm{ }}{\bf{Z}}_k^\textbf{D}{\rm{ }}\underline  \succ  {\rm{ }}0,{\rm{ }}{\bf{Z}}_k^{{\textbf{R}^h}}{\rm{ }}\underline  \succ  {\rm{ }}0,\lambda _k^\textbf{Q} \ge 0,\lambda _k^\textbf{D} \ge 0,\lambda _k^{{\textbf{R}^h}} \ge 0
\nonumber \\
& \textmd{and} \quad {\bf{\textbf{Z}}}_k^\textbf{D}\textmd{ is diagonal matrix}
\label{eqn: Lagrange Function}
\end{align}
%%%%%%%%%%%%%%%%%%%%%%%%%%%%%%%%%%%%%%%%%%%%%%%%%%%%%%%%%%%%%%%%%%%%%%%%%%%%%%%%%%%%%%%%%%%%%%%%%%%%%%%%%%%%%%%%%%%%%%%%%%%%%%%%
where the requirement of being diagonal on the dual variable ${\bf{Z}}_k^\textbf{D}$ comes from the fact
that ${\bf{\Delta }}_k^\textbf{D}$ needs to be diagonal. Since the problem in \eqref{eqn: Lagrange Function} is convex, we can use the dual Lagrange function and Karush-Kuhn-Tucker (KKT) conditions to find the global optimum value of the
problem \cite{Boyd}. Applying the KKT conditions, the Lagrange function will be
\begin{eqnarray}
\lefteqn{L}%\left( {\lambda _k^\textbf{Q},\lambda _k^\textbf{D},\lambda _k^{{\textbf{R}^h}},{\bf{Z}}_k^\textbf{Q},{\bf{Z}}_k^\textbf{D},{\bf{Z}}_k^{{\textbf{R}^h}},{\bf{\Delta }}_k^\textbf{Q},{\bf{\Delta }}_k^\textbf{D},{\bf{\Delta }}_k^{{\textbf{R}^h}}} \right)}
%\nonumber \\
&=&\!\!\!\!\!\!{\textbf{w}^H}\!\!\left( {{\bf{R}}_k^h\!\! - \!{\gamma _k}\left( {{{\bf{Q}}_k}\!\!+\!\!{{\bf{D}}_k}} \right) + {\bf{\Delta}} _k^{{\textbf{R}^h}} - {\gamma _{th}}\left( {{\bf{\Delta}} _k^\textbf{Q} + {\bf{\Delta}} _k^\textbf{D}} \right)} \right)\textbf{w}
\nonumber \\
&& \hspace{-6mm} + {\mkern 1mu} {\mkern 1mu} \lambda _k^\textbf{Q}{\mkern 1mu} \left( {{{\left\| {\left. {{\bf{\Delta}} _k^\textbf{Q}} \right\|} \right.}^2}
- {{\left( {\varepsilon _k^\textbf{Q}} \right)}^2}} \right)
+ \lambda _k^\textbf{D}{\mkern 1mu} \left( {{{\left\| {\left. {{\bf{\Delta}} _k^\textbf{D}} \right\|} \right.}^2} - {{\left( {\varepsilon _k^\textbf{D}} \right)}^2}} \right)
\nonumber \\
&& \hspace{-6mm} + {\mkern 1mu} \lambda _k^{{\textbf{R}^h}}{\mkern 1mu} \left( {{{\left\| {\left. {{\bf{\Delta}} _k^{{\textbf{R}^h}}} \right\|} \right.}^2} - {{\left( {\varepsilon _k^{{\textbf{R}^h}}} \right)}^2}} \right) - Tr\left( {\textbf{Z}_k^\textbf{Q}\left( {{\textbf{Q}_k} + {\bf{\Delta}} _k^\textbf{Q}} \right)} \right)
\nonumber \\
&& \hspace{-6mm} + Tr\left( {\textbf{Z}_k^\textbf{D}{\mkern 1mu} \left( {{\textbf{D}_k}\!\!+\!\!{\bf{\Delta}} _k^\textbf{D}} \right) + {\mkern 1mu} {\mkern 1mu}\!\!Z_k^{{\textbf{R}^h}}\!\!\left( {\textbf{R}_h^k + {\bf{\Delta}} _k^{{\textbf{R}^h}}} \right)} \right) - \sigma _n^2{\gamma _k}
\label{eqn: Lagrange Function 2}
\end{eqnarray}
%%%%%%%%%%%%%%%%%%%%%%%%%%%%%%%%%%%%%%%%%%%%%%%%%%%%%%%%%%%%%%%%%%%%%%%%%%%%%%%%%%%%%%%%%%%%%%%%%%%%%%%%%%%%%%%%%%%%%%%%%%%%%%%%
where $\left\{ {\lambda _k^\textbf{Q}} \right\}_{k = 1}^d$, $\left\{ {\lambda _k^\textbf{D}} \right\}_{k = 1}^d$
and $\left\{ {\lambda _k^{{\textbf{R}^h}}} \right\}_{k = 1}^d$ are the non-negative dual variables. For the sake
of simplicity, we denote the above function by ${L_k}$.
%By introducing ${\bf{X}} = {\bf{w}}{{\bf{w}}^H}$.
%we
%can rewrite \eqref{eqn: Lagrange Function 2} as
Using~\cite{MatrixCookbook} and setting zero the derivatives of $L_k$
with respect to ${\bf{\Delta}} _k^\textbf{Q}$,
${\bf{\Delta}} _k^\textbf{D}$ and
${\bf{\Delta}} _k^{{\textbf{R}^h}}$,
we  have
\begin{align}
{\bf{\Delta}} _k^\textbf{Q} = \frac{{\textbf{Z}_k^\textbf{Q} + {\gamma _{th}}\textbf{X}}}{{2\lambda _k^\textbf{Q}}}, \,
{\bf{\Delta}} _k^\textbf{D} = \frac{{\textbf{Z}_k^\textbf{D} + {\gamma _{th}}\textbf{X}}}{{2\lambda _k^\textbf{D}}} \odot \textbf{I}, \, {\bf{\Delta }} _k^{{\textbf{R}^h}} = \frac{{\textbf{Z}_k^{{\textbf{R}^h}} - \textbf{X}}}{{2{\mkern 1mu} \lambda _k^{{\textbf{R}^h}}}}
\nonumber%\label{eqn: Delta_D_k Derivate answer}
\end{align}
%%%%%%%%%%%%%%%%%%%%%%%%%%%%%%%%%%%%%%%%%%%%%%%%%%%%%%%%%%%%%%%%%%%%%%%%%%%%%%%%%%%%%%%%%%%%%%%%%%%%%%%%%%%%%%%%%%%%%%%%%%%%%%%%
%By solving the equation \eqref{eqn: Delta_D_k Derivate}, the value
%of ${\bf{\Delta}} _k^{{\textbf{R}^h}}$ can be obtained as relation \eqref{eqn: Delta_D_R_h Derivate answer}.
%%%%%%%%%%%%%%%%%%%%%%%%%%%%%%%%%%%%%%%%%%%%%%%%%%%%%%%%%%%%% EQUATION %%%%%%%%%%%%%%%%%%%%%%%%%%%%%%%%%%%%%%%%%%%%%%%%%%%%%%%%%
%\begin{align}
%&\frac{{\partial {{\bf{L}}_k}\,}}{{\partial {\bf{\Delta }}_k^{{\textbf{R}^h}}}} = {\bf{X}} + \,\,2\,\lambda _k^{{\textbf{R}^h}}{\bf{\Delta }}_k^{{\textbf{R}^h}} - {\bf{Z}}_k^{{\textbf{R}^h}} = 0
%\label{eqn: Delta_D_k Derivate}
%\end{align}
%\begin{align}
%&{\bf{\Delta }} _k^{{\textbf{R}^h}} = \frac{{\textbf{Z}_k^{{\textbf{R}^h}} - \textbf{X}}}{{2{\mkern 1mu} \lambda _k^{{\textbf{R}^h}}}}
%\label{eqn: Delta_D_R_h Derivate answer}
%\end{align}
%%%%%%%%%%%%%%%%%%%%%%%%%%%%%%%%%%%%%%%%%%%%%%%%%%%%%%%%%%%%%%%%%%%%%%%%%%%%%%%%%%%%%%%%%%%%%%%%%%%%%%%%%%%%%%%%%%%%%%%%%%%%%%%%
Inserting the derived values
for ${\bf{\Delta}} _k^\textbf{Q}$, ${\bf{\Delta}} _k^\textbf{D}$,
${\bf{\Delta}} _k^{{\textbf{R}^h}}$ in
\eqref{eqn: Lagrange Function 2} and
maximizing the relation with respect
to $\lambda _k^\textbf{Q}$, $\lambda _k^\textbf{D}$ and $\lambda _k^{{\textbf{R}^h}}$, leads to the
following form for the Lagrange dual function after some algebraic manipulations and using ${\bf{X}} = {\bf{w}}{{\bf{w}}^H}$
\begin{subequations}\label{eqn: Dual Lagrange maximized}
\begin{IEEEeqnarray}{lll}
\hspace{-.9 cm}
\mathop {\rm{Maximize} }\limits_{\textbf{Z}_k^\textbf{Q},\textbf{Z}_k^\textbf{D},\textbf{Z}_k^{{\textbf{R}^h}}} && {M\left( {\bf{Z}}_k^\textbf{Q},{\bf{Z}}_k^\textbf{D},{\bf{Z}}_k^{{\textbf{R}^h}},\bf{X} \right)}
\label{eqn: Dual Lagrange maximizeda}\\
\rm{s.\,t.  }&& \,\, \textbf{Z}_k^\textbf{Q}{\rm{ }}\underline  \succ  {\rm{ }}0,\,\,\,\,\textbf{Z}_k^\textbf{D}{\rm{ }}\underline  \succ  {\rm{ }}0,\,\,\,\,\textbf{Z}_k^{{\textbf{R}^h}}{\rm{ }}\underline  \succ  {\rm{ }}0{\rm{      }} \,\,\,\, \forall \,k = 1,...,d
\label{eqn: Dual Lagrange maximizedb}
\end{IEEEeqnarray}
\end{subequations}
where
\begin{IEEEeqnarray}{rcl}
%\vpsace{-1 cm}
%\begin{equation}
 {M\left( \, . \, \right)} &=& Tr\left( {\textbf{X}{\textbf{T}_k}} \right)
-\frac{1}{2}{{\left\| {\left( {\textbf{Z}_k^\textbf{D} + {\gamma _{th}}\textbf{X}} \right) \circ \textbf{I}} \right\|}}\varepsilon _k^\textbf{D}+
\nonumber \\
&&Tr\left( { - \textbf{Z}_k^{{\textbf{R}^h}}\textbf{R}_k^h-\textbf{Z}_k^\textbf{Q}{\textbf{Q}_k} - \textbf{Z}_k^\textbf{D}{\textbf{D}_k} }\right)
-\frac{\varepsilon _k^{{\textbf{R}^h}}}{2}{{\left\| {\textbf{Z}_k^{{\textbf{R}^h}} - \textbf{X}} \right\|}}\nonumber \\
&&
- \frac{\varepsilon _k^\textbf{Q}}{2}{{\left\| {\textbf{Z}_k^\textbf{Q} + {\gamma _{th}}\textbf{X}} \right\|}} -\sigma _n^2{\gamma _k} \nonumber
\end{IEEEeqnarray}
%\begin{IEEEeqnarray}{rcl}
%%\vpsace{-1 cm}
%%\begin{equation}
%
% {M\!\!\left( \, . \, \right)}\!\!=\!\!Tr\!\!\left( {\textbf{X}{\textbf{T}_k}} \right)\!\!
%-\!\!\frac{1}{2}{{\left\| {\left( {\textbf{Z}_k^\textbf{D} + {\gamma _{th}}\textbf{X}} \right) \circ \textbf{I}} \right\|}}\varepsilon _k^\textbf{D}+
%\nonumber \\
%\!\!&&\!Tr\!\!\left( \!\!{ - \textbf{Z}_k^{{\textbf{R}^h}}\!\!\textbf{R}_k^h\!\!-\!\!\textbf{Z}_k^\textbf{Q}{\textbf{Q}_k}\!\! \!\!-\!\! \textbf{Z}_k^\textbf{D}{\textbf{D}_k} }\!\!\right)
%\!\!-\!\!\frac{\varepsilon _k^{{\textbf{R}^h}}}{2}{{\left\| {\textbf{Z}_k^{{\textbf{R}^h}}\!\!\!\! -\!\! \textbf{X}} \right\|}}
%\!\!-\!\! \frac{\varepsilon _k^\textbf{Q}}{2}\!{{\left\| {\textbf{Z}_k^\textbf{Q}\!\! + \!\!{\gamma _{th}}\textbf{X}} \right\|}}\!\! -\!\!\sigma _n^2{\gamma _k} \nonumber
%\end{IEEEeqnarray}
%%%%%%%%%%%%%%%%%%%%%%%%%%%%%%%%%%%%%%%%%%%%%%%%%%%%%%%%%%%%%%%%%%%%%%%%%%%%%%%%%%%%%%%%%%%%%%%%%%%%%%%%%%%%%%%%%%%%%%%%%%%%%%%%
By substituting~\eqref{eqn: Dual Lagrange maximized}
and~\eqref{eqn: Robust Optimization Objective 3}
in~\eqref{eqn: Robust Optimization Problem}, we can
write our main robust optimization problem as
%%%%%%%%%%%%%%%%%%%%%%%%%%%%%%%%%%%%%%%%%%%%%%%%%%%%%%%%%%%%% EQUATION %%%%%%%%%%%%%%%%%%%%%%%%%%%%%%%%%%%%%%%%%%%%%%%%%%%%%%%%%
\begin{IEEEeqnarray}{lCl}
\mathop {{\rm{min.}}}\limits_\textbf{X} &\quad& Tr\left( {\textbf{X}\left( {\textbf{D} + {\varepsilon _\textbf{D}}\textbf{I}} \right)} \right)
\nonumber \\
\rm{s.\,t. \!\! }  &&  \mathop {\max }\limits_{\textbf{Z}_k^\textbf{Q},\textbf{Z}_k^\textbf{D},
\textbf{Z}_k^{{\textbf{R}^h}}} \,\, {M \left( {\bf{Z}}_k^\textbf{Q},{\bf{Z}}_k^\textbf{D},{\bf{Z}}_k^{{\textbf{R}^h}} \right)}\,\,
\geq0\nonumber\\
&& {\textmd{Constraints}}~\eqref{eqn: Dual Lagrange maximizedb}
%\bigg(Tr\left( {\textbf{X}{\textbf{T}_k}} \right) - \sigma _n^2{\gamma _k}
%\nonumber \\
%&& + Tr\left( { - \textbf{Z}_k^{{\textbf{R}^h}}\textbf{R}_k^h - \textbf{Z}_k^\textbf{Q}{\textbf{Q}_k} - \textbf{Z}_k^\textbf{D}{\textbf{D}_k}} \right)
%\nonumber \\
%&&- \frac{{\left\| {\textbf{Z}_k^{{\textbf{R}^h}} - \textbf{X}} \right\|}}{2}\varepsilon _k^{{\textbf{R}^h}} - \frac{{\left\| {\textbf{Z}_k^\textbf{Q} + {\gamma _{th}}\textbf{X}} \right\|}}{2}\varepsilon _k^\textbf{Q}
%\nonumber\\
%&&- \frac{{\left\| {\left( {\textbf{Z}_k^\textbf{D} + {\gamma _{th}}\textbf{X}} \right) \circ \textbf{I}} \right\|}}{2}\varepsilon _k^\textbf{D}\bigg)\geq \, 0
%\nonumber \\
%&& \,\,\,\, \textbf{Z}_k^\textbf{Q}{\rm{ }}\underline  \succ  {\rm{ }}0,\,\,\,\,\textbf{Z}_k^\textbf{D}{\rm{ }}\underline  \succ  {\rm{ }}0,\,\,\,\,\textbf{Z}_k^{{\textbf{R}^h}}{\rm{ }}\underline  \succ  {\rm{ }}0{\rm{      }} \,\,\,\, \forall \,\,\,\,k = 1,...,d.
,\,{\textmd{and}\,\,\rm{Rank}}\left( \textbf{X} \right){\rm{ = 1}}
\label{eqn: Robust final}
\end{IEEEeqnarray}
%%%%%%%%%%%%%%%%%%%%%%%%%%%%%%%%%%%%%%%%%%%%%%%%%%%%%%%%%%%%%%%%%%%%%%%%%%%%%%%%%%%%%%%%%%%%%%%%%%%%%%%%%%%%%%%%%%%%%%%%%%%%%%%%
Consider the fact that maximum of the expression $Tr\left( { - \textbf{Z}_k^{{\textbf{R}^h}}\textbf{R}_k^h -\textbf{Z}_k^\textbf{Q}{\textbf{Q}_k} - \textbf{Z}_k^\textbf{D}{\textbf{D}_k}} \right)$ is equal to
zero (since $\textbf{Z}_k^\textbf{Q}{\rm{ }}\underline  \succ  {\rm{ }}0$, $\textbf{Z}_k^\textbf{D}{\rm{ }}\underline  \succ  {\rm{ }}0$, $\textbf{Z}_k^{{\textbf{R}^h}}{\rm{ }}\underline  \succ  {\rm{ }}0{\rm{      }}$, $\forall \,\,\,\,k = 1,...,d.$) and setting zero all the norm bounds in \eqref{eqn: Robust final}, the problem
in \eqref{eqn: Robust final} is transformed to the non-robust problem stated in~\cite{Fazeli-Dehkordy:2009:Trans}. If all the constraints in \eqref{eqn: Robust final}  be above zero, then the maximum constraint
will also satisfy the inequality, so \eqref{eqn: Robust final} can be
simplified as follows
%%%%%%%%%%%%%%%%%%%%%%%%%%%%%%%%%%%%%%%%%%%%%%%%%%%%%%%%%%%%% EQUATION %%%%%%%%%%%%%%%%%%%%%%%%%%%%%%%%%%%%%%%%%%%%%%%%%%%%%%%%%
\begin{IEEEeqnarray}{lCl}
\mathop {{\rm{min.}}}\limits_\textbf{X} \,\,\,&&Tr\left( {\textbf{X}\left( {\textbf{D} + {\varepsilon _\textbf{D}}\textbf{I}} \right)} \right)
\nonumber \\
\rm{s.\,t.  }&& \,\, {M \left( {\bf{Z}}_k^\textbf{Q},{\bf{Z}}_k^\textbf{D},{\bf{Z}}_k^{{\textbf{R}^h}} \right)}\geq0\quad\forall k=1,\cdots,d,
\nonumber\\
&& {\textmd{Constraints}}~\eqref{eqn: Dual Lagrange maximizedb},\,{\textmd{and}\,\,\rm{Rank}}\left( \textbf{X} \right){\rm{ = 1}}
%Tr\left( {\textbf{X}{\textbf{T}_k}} \right) - \sigma _n^2{\gamma _k} + Tr\left( { - \textbf{Z}_k^{{\textbf{R}^h}}\textbf{R}_k^h - \textbf{Z}_k^\textbf{Q}{\textbf{Q}_k} - \textbf{Z}_k^\textbf{D}{\textbf{D}_k}} \right)
%\nonumber \\
%&& - \frac{{\left\| {\textbf{Z}_k^{{\textbf{R}^h}} - \textbf{X}} \right\|}}{2}\varepsilon _k^{{\textbf{R}^h}} - \frac{{\left\| {\textbf{Z}_k^\textbf{Q} + {\gamma _{th}}\textbf{X}} \right\|}}{2}\varepsilon _k^\textbf{Q}
%\nonumber \\
%&&- \frac{{\left\| {\left( {\textbf{Z}_k^\textbf{D} + {\gamma _{th}}\textbf{X}} \right) \circ \textbf{I}} \right\|}}{2}\varepsilon _k^\textbf{D}\geq0
%\nonumber \\
%&& \,\,\,\, \textbf{Z}_k^\textbf{Q}{\rm{ }}\underline  \succ  {\rm{ }}0,\,\,\,\,\textbf{Z}_k^\textbf{D}{\rm{ }}\underline  \succ  {\rm{ }}0,\,\,\,\,\textbf{Z}_k^{{\textbf{R}^h}}{\rm{ }}\underline  \succ  {\rm{ }}0{\rm{      }} \,\,\,\, \forall \,\,\,\,k = 1,...,d.
\label{eqn: Robust final 1}
\end{IEEEeqnarray}
%%%%%%%%%%%%%%%%%%%%%%%%%%%%%%%%%%%%%%%%%%%%%%%%%%%%%%%%%%%%%%%%%%%%%%%%%%%%%%%%%%%%%%%%%%%%%%%%%%%%%%%%%%%%%%%%%%%%%%%%%%%%%%%%
Note that the objective function in \eqref{eqn: Robust final 1} is linear and all the constraints
except the last one are conic convex. We drop this non-convex constraint to relax the problem into a
convex optimization problem. The well known semi-definite problem (SDP) solvers such as SeDuMi or CVX can be used for solving the above problem
by semi-definite programming
in polynomial time using interior point methods.
Since the solution of \eqref{eqn: Robust final 1} is not always rank one, randomization techniques \cite{Fazeli-Dehkordy:2009:Trans} can be used to obtain an approximate solution of the original problem from the solution of the relaxed problem.
However, our simulation results show that the rank of $\textbf{X}$ is always one when $d<3$. This has been also reported in \cite{Zhiquan:2007} analytically for $d<3$. If the optimum solution is a rank-one matrix, the principal value of the matrix is used to determine $\textbf{w}$, otherwise the best rank one approximation is obtained using the procedure described in \cite{Sidiropoulos:2006:Transactions}.
Note that the minimum value of the relaxed form (without rank constraint) of the
problem in \eqref{eqn: Robust final 1} is a lower bound for the minimum value of the original
problem in \eqref{eqn: Robust final 1}.
% In the other words, the gap between the original and the relaxed
%version of the problem in \eqref{eqn: Robust final 1} is the duality gap.
As mentioned after \eqref{eqn: Perturbation Limit}, the optimal solution of \eqref{eqn: Robust final 1} is still suboptimal solution for \eqref{eqn:original_SIP} as it happens for \cite{Chalise:2010:Trans,chaliseMIMO1} too, but if uncertainties of the perturbation matrices are occurred due to the quantization noise of related matrices, they can be assumed independent and both problems have the same optimal solutions.
%Note that, the semi-definite
%relaxation provides the same lower bound to the original problem as the dual problem does. Interest
%readers can see \cite{Fazeli-Dehkordy:2009:Trans} to follow the proof of this claim.
%%%%%%%%%%%%%%%%%%%%%%%%%%%%%%%%%%%%%%%%%%%%%%%%%%%%%%%%%%%% SECTION %%%%%%%%%%%%%%%%%%%%%%%%%%%%%%%%%%%%%%%%%%%%%%%%%%%%%%%%%
\section{Simulation and Numerical Results}
\label{Sec: Simulation}
%%%%%%%%%%%%%%%%%%%%%%%%%%%%%%%%%%%%%%%%%%%%%%%%%%%%%%%%%%%% SECTION %%%%%%%%%%%%%%%%%%%%%%%%%%%%%%%%%%%%%%%%%%%%%%%%%%%%%%%%%
In this section, we use  Monte-Carlo simulations to compare our accurate robust with MoM robust\cite{Chalise:2010:Trans} and non-robust power allocation methods. In all simulations, we assume $15$ relays ($R=15$) and $2$ users ($d=2$).
 The noise power and source transmit power are assumed to be $0 dB$ Watt and the channels are realized randomly with Rayleigh fading distribution. For the robust case, we suppose that the maximum perturbation norm of $\textbf{D}$, ${\textbf{D}_k}$, $\textbf{R}_k^h$ and $\textbf{Q}_k$ matrices are $1\%$ of their
Frobenius norms which we denote $\varepsilon  = 0.01{\left\| {\,.\,} \right\|_F}$ by meaning that ${\varepsilon _{\bf{X}}} = 0.01{\left\| {\bf{X}} \right\|_F}$   for perturbation bound of any perturbed matrix $\mathbf{X}$.
%for example, $\varepsilon _k^{{\textbf{R}^h}} = \left\| {{\bf{\Delta}} _k^{{R^h}}} \right\| = 0.05\left\| {\textbf{R}_k^h} \right\|$.
%The elements of ${\bf{\Delta _D}}$ are taken as independent identically
%zero-mean random uniform variables
%with matrix bounded norm are scattered uniformly between $0$ and $\varepsilon$.
%Three simulation scenarios are brought up to compare the non-robust and robust methods.

\begin{figure}
\centering
\hspace{-1 cm}
\includegraphics[width=9.5cm]{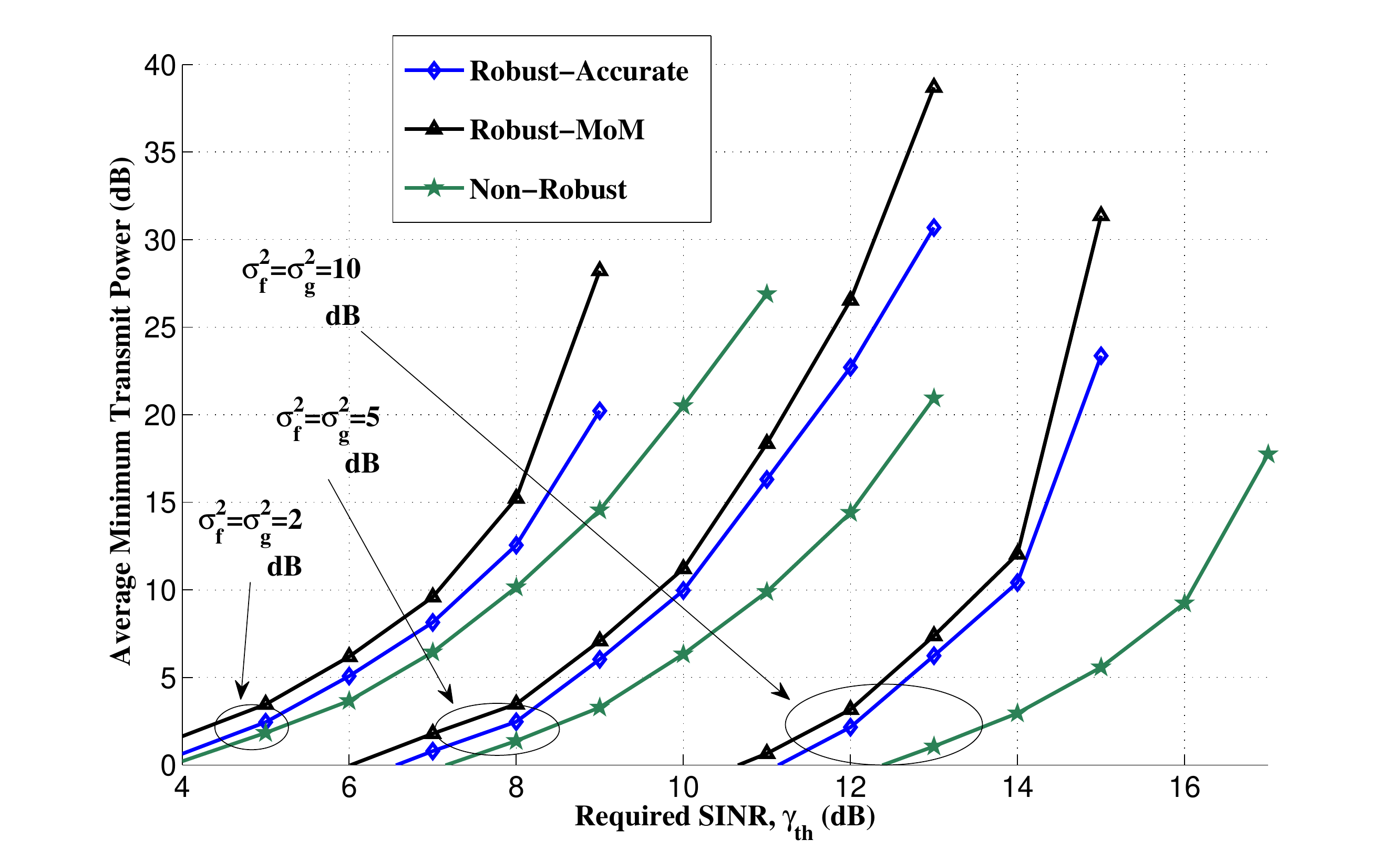}
\vspace{-5mm} \caption{Average minimum transmit power versus the required SINR for $\varepsilon  = 1\% {\left\| . \right\|_F}$ perturbation}
\label{fig: Power 1 Percent}
\vspace{-3mm}
\end{figure}

In the first simulation setup, the average transmitted powers for the non-robust and both of the robust methods (MoM and Accurate) are depicted in Fig.~\ref{fig: Power 1 Percent} with respect to the required received SINR for different channel variances. Figure \ref{fig: Power 1 Percent}
shows the extra amount of the power that the robust methods consume as the expense of resistance against channel perturbations, while holding the QoS constraint in its desired range. It can be seen that our proposed accurate robust method outperforms MoM robust method for the difference channel variances.

%In the second simulation setup, the histogram of one of the user's received SINR is drawn in Fig.~\ref{fig: Hist 1 Percent} for a channel variance $5$dB and $40000$ ensembles of channel realizations. This scenario shows the vulnerability of non-robust method to even small amount of channel uncertainty and also the resistance of our proposed robust method.
%Fig~.\ref{fig: Hist 5 Percent} shows the QoS histogram for both
%non-robust and robust methods when the required SINR of $2$ dB is
%required at the destination. The
%vertical axis shows the number of samples and the horizontal
%axis shows the SINR which these samples
%have provided at the destination.
%It can be seen in Fig.~\ref{fig: Power 5 Percent} that for the required SINR
%of $2$ dB at the destination, the robust method consumes $2$ dB more power than
%the non-robust method; however, about $42\%$ of the samples in the non-robust
%case do not satisfy the QoS constraints while in all the ensembles of the robust
%case, the received SINR is above the required SINR. In other words, for the
%robust method, the outage probability is zero and the QoS condition is
%completely qualified at the destinations.
%$$$$$$$$$$$$$$$$$$$$$$$$$$$$$$$$$$$$$$$$$$$$$$$$$$$$$$$$ FIGURE 3 $$$$$$$$$$$$$$$$$$$$$$$$$$$$$$$$$$$$$$$$$$$$$$$$$$$$$$$
\begin{figure}
\centering
\hspace{-.5 cm}
\includegraphics[width=9.25cm]{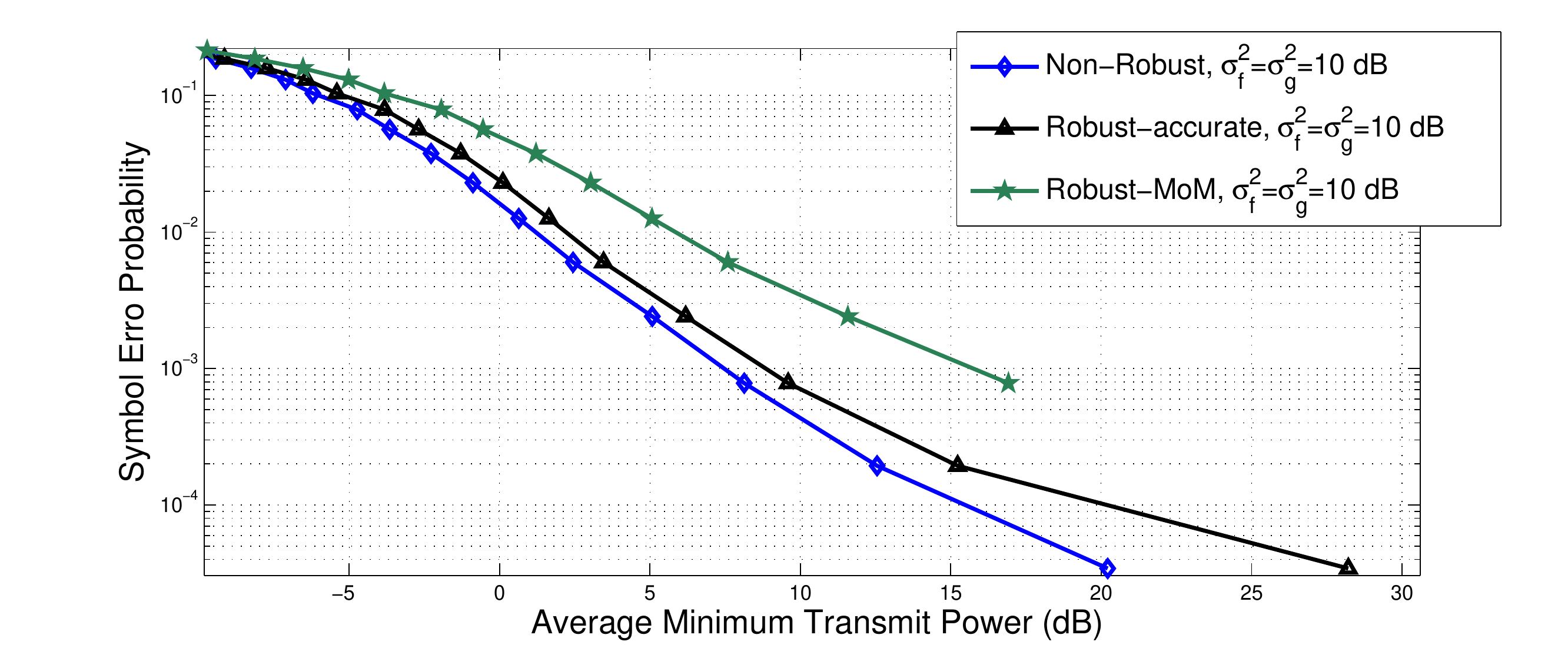}
\vspace{-3mm} \caption{Comparison of worst-case symbol error probability between non-robust and robust systems by sweeping $\gamma_{th}$.}
\label{fig: SEP}
\vspace{-3mm}
\end{figure}

%$$$$$$$$$$$$$$$$$$$$$$$$$$$$$$$$$$$$$$$$$$$$$$$$$$$$$$$$$$$$$$$$$$$$$$$$$$$$$$$$$$$$$$$$$$$$$$$$$$$$$$$$$$$$$$$$$$$$$$$$$
 %For the desired SINR of $\gamma_{th}=2$ dB at the destination, about $32\%$ of the samples do not satisfy the required QoS at the destination, %According to Fig.~\ref{fig: Power 1 Percent}, while the cost of guarantying the QoS for robust method is less than $0.4$ dB with respect to the non-robust method.
In the last scenario, the efficiency of our proposed accurate robust method is compared with MoM robust and non-robust methods  in terms of
symbol error probability (SEP) and average transmitted power. Figure \ref{fig: SEP} shows the SEP versus average transmit power of the relays for the same variance
$\sigma_f^2=\sigma_g^2=10\textrm{dB}$ of the elements of the channels vector $\mathbf{f, g}$ over $3000$ iterations and $1\%$ perturbation while at least 20\% of the realization are feasible. Please note that the non-robust method is evaluated in non-perturbed settings to draw a benchmark for computing the power loss of other two robust methods. It is observed that for a specific
amount of SEP, the accurate robust design outperforms the MoM and non-robust designs in terms of power consumption.
%$$$$$$$$$$$$$$$$$$$$$$$$$$$$$$$$$$$$$$$$$$$$$$$$$$$$$$$$$$$$$$$$$$$$$$$$$$$$$$$$$$$$$$$$$$$$$$$$$$$$$$$$$$$$$$$$$$$$$$$$$
%$$$$$$$$$$$$$$$$$$$$$$$$$$$$$$$$$$$$$$$$$$$$$$$$$$$$$$$$$$$$$$$$$$$$$$$$$$$$$$$$$$$$$$$$$$$$$$$$$$$$$$$$$$$$$$$$$$$$$$$$$
%%%%%%%%%%%%%%%%%%%%%%%%%%%%%%%%%%%%%%%%%%%%%%%%%%%%%%%%%%%% SECTION %%%%%%%%%%%%%%%%%%%%%%%%%%%%%%%%%%%%%%%%%%%%%%%%%%%%%%%%%
%\vspace{-3mm}
\section*{Acknowledgment}
This works is done in  SSWC-Lab of K. N. Toosi University of Technology and financially supported by Iranian Research Institute for ICT recently renamed to CyberSpace research institute.
%\section*{Acknowledgment}
%This works is done in  SSWC-Lab of KNTU and supported by Iranian Research Institute for ICT.
\section{Conclusion}
\label{Sec: Conclusion}
%%%%%%%%%%%%%%%%%%%%%%%%%%%%%%%%%%%%%%%%%%%%%%%%%%%%%%%%%%%% SECTION %%%%%%%%%%%%%%%%%%%%%%%%%%%%%%%%%%%%%%%%%%%%%%%%%%%%%%%%%
In this paper, the robust optimal power allocation algorithm for multi-user cooperative networks was solved in a more accurate approach compared to the previous works. The proposed approach assumed uncertaintiy on all channels in the QoS aware beamforming problem to perform a robust design. The simulation results have shown the superiority of the proposed robust method compared to the previous robust and non-robust methods.
%%%%%%%%%%%%%%%%%%%%%%%%%%%%%%%%%%%%%%%%%%%%%%%%%%%%%%%%%%%% SECTION %%%%%%%%%%%%%%%%%%%%%%%%%%%%%%%%%%%%%%%%%%%%%%%%%%%%%%%%%
\vspace{-2mm}
%%%%%%%%%%%%%%%%%%%%%%%%%%%%%%%%%%%%%%%%%%%%%%%%%%%%%%%%%%%% SECTION %%%%%%%%%%%%%%%%%%%%%%%%%%%%%%%%%%%%%%%%%%%%%%%%%%%%%%%%%

%%%%%%%%%%%%%%%%%%%%%%%%%%%%%%%%%%%%%%%%%%%%%%%%%%%%%%%%%%%% SECTION %%%%%%%%%%%%%%%%%%%%%%%%%%%%%%%%%%%%%%%%%%%%%%%%%%%%%%%%%
%\bibliographystyle{IEEEtran}
%\bibliography{refrence}
\end{document}